\documentclass[]{article}

\usepackage{epsfig,amsmath}
\usepackage{graphicx}

\begin{document}

%
%
\title{Catalytic Reaction Sets, Decay, and the Preservation of Information}
\author{Wim Hordijk and Jos\'e F. Fontanari}
\date{}

\maketitle

%
%
\begin{abstract}
We study the ability to maintain information in a population of reacting polymers under the
influence of decay, i.e., spontaneous breakdown of large polymers. At a certain decay rate,
it becomes impossible to maintain a significant concentration of large polymers, while it is
still possible to maintain sets of smaller polymers that can maintain the same amount of
information. We use a genetic algorithm to evolve reaction sets to generate specific polymer
distributions under the influence of decay. In these evolved reaction sets, the beginnings of
hypercycle-type structures can be observed, which are believed to have been an important step
toward the evolution of the first living cells.
\end{abstract}

%
%
\section{Introduction} \label{sec:Introduction}

The amount of information that can be stored and preserved in a population of reacting polymers
depends crucially on the reaction efficiencies. For example, larger polymers are more likely to
break down into smaller parts than shorter polymers. So, to maintain a significant concentration
of a certain large polymer, there have to exist highly efficient reactions building up these large
polymers from smaller ones. In fact, the efficiencies (or reaction rates) of these reactions have
to be larger than the rate of polymer breakdown. This is somewhat equivalent to the error threshold
phenomenon in self-replicating polynucleotides, where the amount of information that can be
preserved is limited by the replication accuracy.

One proposed solution to circumvent the error catastrophe is the {\em hypercycle}
\cite{Eigen:77,Eigen:79}, a catalytic feedback loop where each polymer increases the efficiency of
the creation of the next polymer in a (closed) reaction loop. This way, parts of the information
can be stored in smaller polymers that help each other in maintaining a large enough concentration
of each of them. So, besides storing part of the information, such a polymer also acts as a catalyst
in creating one or more other polymers that store other parts of the information. This way, the set
as a whole can preserve the complete information, whereas one large polymer could not.

In this paper, we study the problem of information preservation in a simple model of reacting
polymers. We use a genetic algorithm to evolve catalytic reaction sets to generate a certain target
distribution of polymers under the absence or existence of polymer decay (i.e., breakdown of large
polymers). Our model is based on that in \cite{Lohn:98}, but has some additional features and a
more realistic method of simulating polymer reactions. We then look at the differences between a
target consisting of one large polymer and a target of three smaller ones, the lengths of which add
up to that of the larger one, and if (or how) each target can be reached and maintained under both
the absence and the existence of decay. Such studies have direct relevance to, for example, the
origin of life problem, where it is believed that simple hypercycle-type structures were the
precursor to the first living cells (see, e.g., \cite{MaynardSmith:79}).

In the next section, a method for simulating simple chemical reactions on a computer is reviewed.
In section \ref{sec:EvolChemReacSets} the model for evolving chemical reaction sets is explained.
Section \ref{sec:Results} then presents the results of this model comparing different target
polymer distributions under the absence or existence of decay. Finally, section
\ref{sec:Conclusions} summarizes the main conclusions.

%
%
\section{Simulating Chemical Reactions} \label{sec:SimChemReac}

The model we use considers simple polymers made up of only one type of molecule, and the types of
interactions that are possible are bonding and breaking. The main characteristic of a polymer is
its length, or the number of molecules in the polymer chain. Polymers of length $i$ are denoted
$P_i$. We restrict the length of polymers to a maximum of 35. The bonding reaction simply ``glues''
two polymers of lengths $i$ and $j$ together into one polymer of length $i+j$ (provided that
$i+j \leq 35$). The breaking reaction takes a polymer of length $k$ and splits it into two polymers
of lengths $i$ and $j$ where $i+j = k$. However, only catalytic reactions are considered. In other
words, a reaction can only happen under the influence of an additional polymer that catalyzes the
reaction but which is not involved in the reaction otherwise. A catalyzed bonding reaction is
written as $P_i + P_j + P_k \rightarrow P_{i+j} + P_k$, where $i+j \leq 35$. The catalyst $P_k$ is
not involved in the reaction itself, so it appears again on the right side as a reaction product.
A catalyzed breaking reaction is written as $P_{i+j} + P_k \rightarrow P_i + P_j + P_k$, again with
$i+j \leq 35$ and $P_k$ being the catalyst.

Now suppose we have a reactor with a large number of polymers that is well stirred. Reactions
between polymers happen in this reactor based on the concentrations of the reactants (and
catalysts) in the reactor. Usually, such a system is modeled with a set of coupled ordinary
differential equations (ODE's), one equation for each type of polymer. However, such a system of
ODE's quickly becomes analytically unsolvable or numerically cumbersome. In
\cite{Gillespie:76,Gillespie:77}, a method for numerically simulating such chemical reactions
using a stochastic method is introduced. Instead of calculating changes in polymer concentrations
over very small time steps (the ODE approach), this stochastic method is based on deriving a
reaction probability density function (pdf) $P(\tau, \mu)d\tau$ = probability at time $t$ that the
{\em next} reaction in the reactor will occur in the time interval $(t+\tau, t+\tau+d\tau)$
{\em and} will be of type $\mu$ (given a certain number $M$ of possible reactions). This pdf has
certain parameters, the values of which depend on the current polymer concentrations in the
reactor. The method then uses a Monte Carlo procedure to generate a stream of random numbers that
are interpreted as reaction times and types, and the parameter values of the pdf are updated after
every reaction to reflect the new polymer concentrations. In this simulation method, there is also
a parameter $c_i$, the reaction efficiency, for each of the $M$ reactions. In our model, we use the
same value for $c_i$ for each reaction (i.e., there is no difference in efficiencies for the
different reaction types).For a complete overview of the derivation of the reaction pdf, see
\cite{Gillespie:76,Gillespie:77}.

So, to summarize, we have a set of $N=35$ polymer types $P_i$, and a set of $M$ possible reactions
$R_i$ where each $R_i$ is a catalyzed bonding or breaking reaction. The algorithm for simulating
this polymer reaction system is then:
\begin{enumerate}
\item Set the current time $t=0$, generate an initial polymer type distribution, and calculate the
reaction pdf parameters based on this initial distribution. Set a `''stopping'' time $T$.
\item Generate a random pair $(\tau, \mu)$ from the reaction pdf $P(\tau, \mu)$.
\item Set the time $t = t+\tau$ and perform reaction $R_{\mu}$. Update the polymer type
concentrations and the reaction pdf parameters according to $R_{\mu}$.
\item If $t \geq T$, stop. Otherwise, go to step 2.
\end{enumerate}

%
%
\section{Evolving Chemical Reaction Sets}   \label{sec:EvolChemReacSets}

In \cite{Lohn:98}, a model for evolving catalytic reaction sets was introduced. A genetic algorithm
(GA) \cite{Holland:75,Goldberg:89,Mitchell:96}) was used to evolve a population of reaction sets to
try to find reaction sets that could produce a prespecified polymer distribution given some initial
distribution. Related types of models were used in, for example, \cite{Kauffman:86,Farmer:86}. Here,
a similar approach is used.

The population in our GA consists of reaction sets ${\cal R}$, where each reaction set contains
100 reactions $R_i$ which are catalyzed bonding or breaking reactions. So, each $R_i$ is of the
form $P_i + P_j + P_k \rightarrow P_{i+j} + P_k$ with $i+j \leq 35$ (bonding), or
$P_{i+j} + P_k \rightarrow P_i + P_j + P_k$ with $i+j \leq 35$ (breaking). We use a population
size $S = 100$, and the initial population is created at random, where the fraction of breaking
reactions in each reaction set is a parameter of the algorithm (usually set to 0.2).

The genetic operators are implemented as follows. In crossover, two ``parent'' reaction sets
${\cal R}_{p1}$ and ${\cal R}_{p2}$ are taken from the mating pool, and a random number $c$ (the
crossover point) between 1 and 100 is drawn from the uniform distribution. The first child,
${\cal R}_{c1}$, is then formed by combining the first $c$ reactions $R_i$ from the first parent,
${\cal R}_{p1}$, with the last $100-c$ reactions from the second parent, ${\cal R}_{p2}$. The
second child, ${\cal R}_{c2}$, is formed in a similar way but with the opposite parts of the
parents. The mutation operator simply replaces a reaction $R_i$ in a reaction set with a randomly
chosen new reaction (independent of the reaction being replaced).

For selection, the standard roulette wheel selection method is used.

The fitness function of the GA is implemented as follows. Given an individual ${\cal R}$ from the
GA population and an initial polymer distribution, use the stochastic simulation method as
described in the previous section to iterate this reaction set for $T$ time units (in most runs we
used $T=100$, and the values for the reaction efficiencies were set so that, at least initially,
there are about 100 reactions performed in one time unit). Continue iterating the stochastic
simulation method for another $T$ time units, but after each time unit calculate a ``target value''
$v_t$. At the end, take the average of all target values and return that as the fitness value,
i.e., the fitness of a reaction set ${\cal R}$ is $f_{\cal R} = \sum_{t=1}^T v_t/T$.

In our experiments, we used two different ways of calculating the target values $v_t$. The first
one is $v_t = n_{35}$, which simply means the number of polymers of length 35 (the maximum length)
in the polymer population at time step $t$. So, for this target, the fitness of an individual is
the number of polymers of maximum length, averaged over the second set of $T$ time units. The
second way of measuring the target values is $v_t = n_{10} + n_{12} + n_{13} - \left | n_{10} - 
n_{12} \right | - \left | n_{12} - n_{13} \right |$. So, for this target we try to get as many
polymers of lengths 10, 12, and 13, but in roughly equal numbers (and again averaged over the
second set of time steps). Note that the lengths of these polymers add up to 35, and indeed the
main idea behind this target is to try to get the same ``information'' as in the first target, but
split up in smaller pieces.

Finally, an element of spontaneous polymer breakdown, or decay, is added. In the stochastic
simulation method, next to the set ${\cal R}$ of reactions that forms an individual in the GA
population, there is an independent set of decay reactions $P_{i+j} \rightarrow P_i + P_j$ which
are not catalyzed. The reaction efficiencies of these reactions depend on the length of the
polymer that is being broken down. In our simulations, the efficiency of a decay reaction is some
constant $d$ times the square of the length of the polymer ($i+j$). The constant $d$ is another
parameter in the GA, and can be set to 0 to turn decay off completely, or increased in value for
increasing decay rates.

With this model setup, we can now study the differences between the two different targets (polymers
of maximum length 35, or polymers of lengths 10, 12, and 13) under the influence (or absence) of
decay.

%
%
\section{Results} \label{sec:Results}

\begin{figure}
  \begin{center}
  \epsfig{file=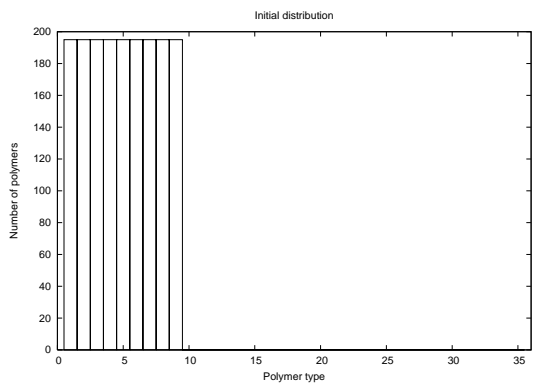, width=2in, angle=270}
  \end{center}
  \caption{The initial polymer distribution. There are 195 polymers for each of the polymer types
           1 to 9, and 0 polymers of each other type.}
  \label{fig:initpop}
\end{figure}

Several GA runs were performed using the two different targets, both with and without decay. In
this section, the main results of these different runs are presented. In the fitness calculations,
an initial polymer distribution as shown in figure \ref{fig:initpop} was used. In this initial
distribution, there are 195 polymers each for the polymer types (lengths) 1 to 9, and 0 polymers of
any other length. So, for both targets, there do not yet exist any target polymers in the initial
polymer population.

\begin{figure}
  \begin{center}
  \begin{tabular}{cc}
  \epsfig{file=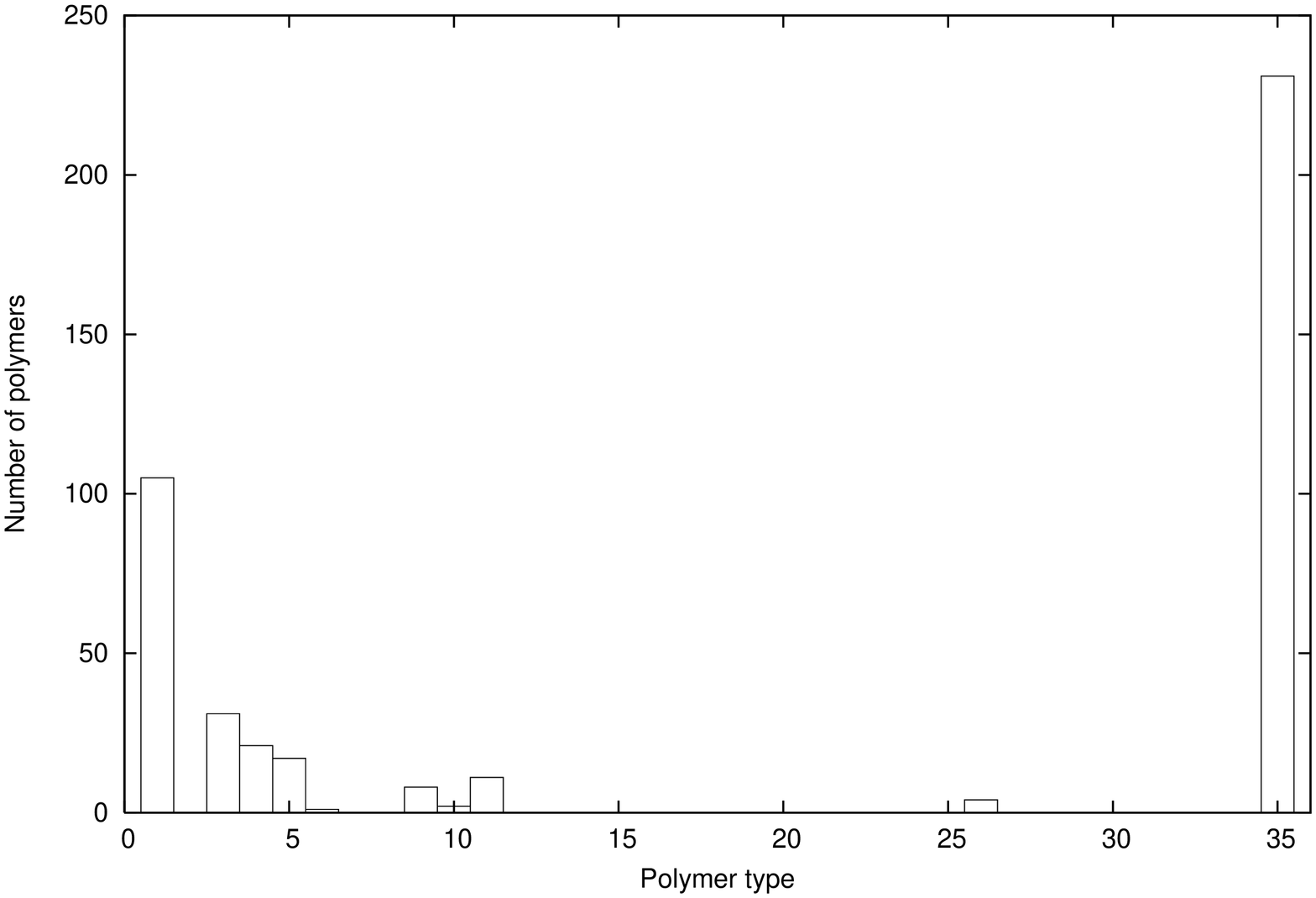, width=2in, angle=270} &
  \epsfig{file=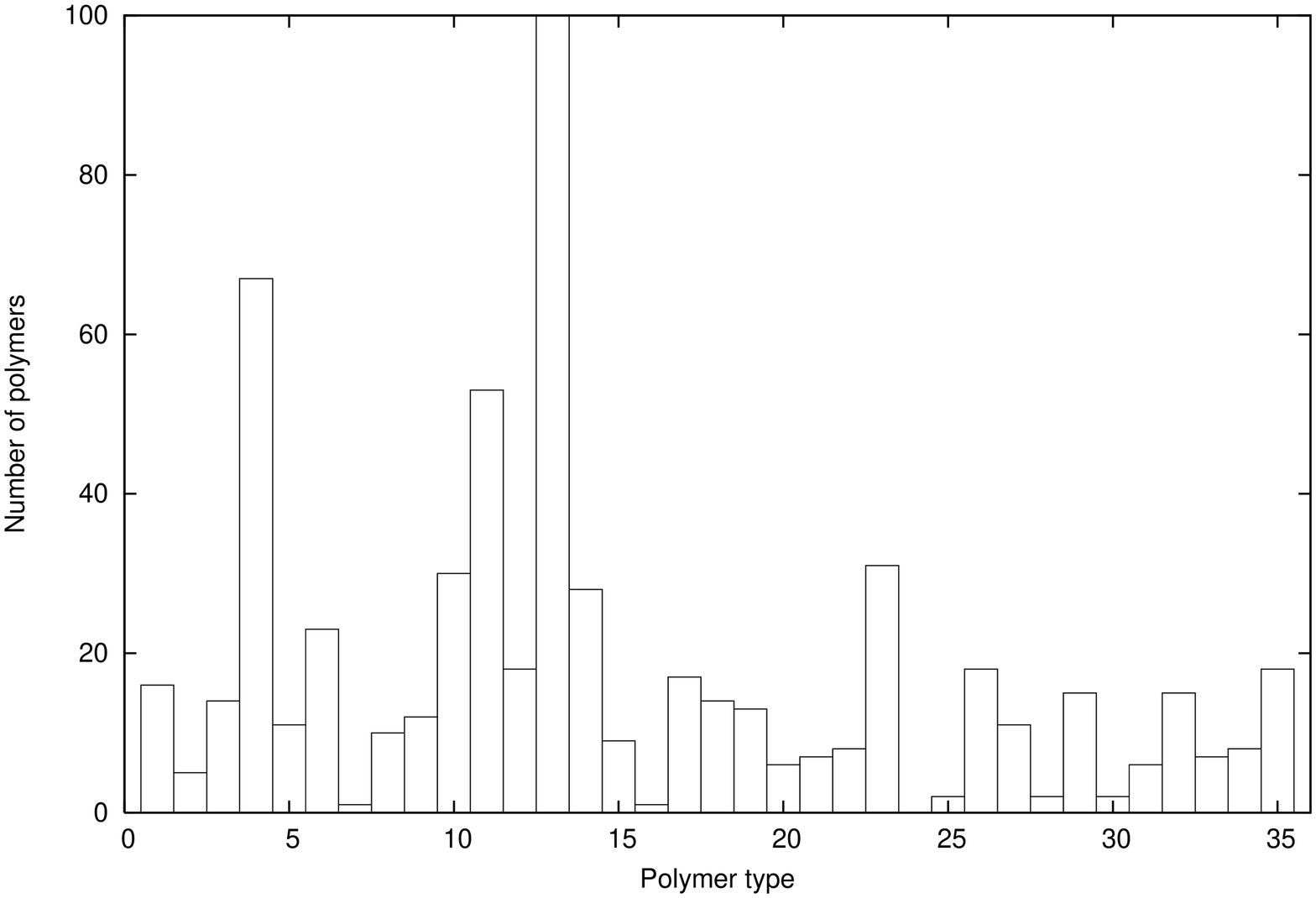, width=2in, angle=270}
  \end{tabular}
  \end{center}
  \caption{Left: The polymer distribution after 1000 time steps for an evolved reaction set for the
           length 35 target with no decay. This reaction set produces around 230 polymers of
           the maximum length (35). Right: The result of the same reaction set, but this time
           iterated with decay at $d=0.0001$. It only manages to maintain around 20 polymers of
           length 35.}
  \label{fig01}
\end{figure}

First, the GA was run on the first target (polymers of length 35) without any decay (i.e., the
decay parameter $d$ was set to 0.0). In every run the GA was able to find reaction sets that
produce around 200 polymers of this length. Figure \ref{fig01} (left) shows a typical result. Even
though in the GA runs the reaction simulations were run up to $T=200$, all the results shown in this
sections are for $T=1000$. For example, in figure \ref{fig01}, one of the best individuals found
by the GA was taken, and this reaction set was then re-iterated for $T=1000$ time steps (starting
with the same initial distribution as shown in figure \ref{fig:initpop}) to make sure that some
equilibrium has been reached. As the figure shows, this reaction set produces slightly more than
200 polymers of length 35.

Although this reaction set was evolved without using decay, we can ask how it performs when
iterated with decay turned on. Figure \ref{fig01} (right) shows the equilibrium distribution (again
at $T=1000$) of the same reaction set, but with the decay parameter set to $d=0.0001$. In this case,
it produces less than 20 polymers of maximum length, more than 10 times less compared to the
no-decay case.

\begin{figure}
  \begin{center}
  \epsfig{file=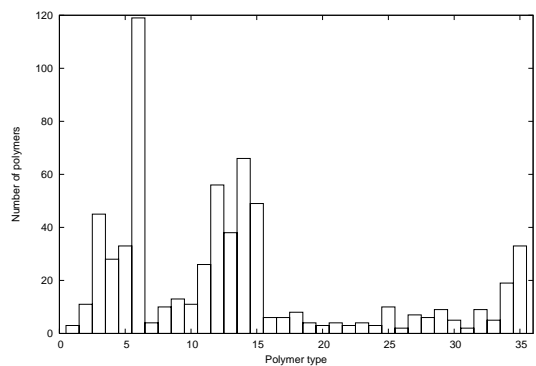, width=2in, angle=270}
  \end{center}
  \caption{The result of one of the best reaction sets evolved with decay at $d=0.0001$, for the
           maximum length target. There are only around 35 target polymers.}
  \label{fig03}
\end{figure}

Of course this reaction set was not evolved to deal with decay, and so it is expected to perform
poorly under the influence of decay. Next, the GA was run again on the same target, but this time
with decay (again with $d=0.0001$). Figure \ref{fig03} shows the equilibrium distribution of one
of the best reaction sets found by the GA in this case. As the figure shows, even though the
reaction set was evolved under the influence of decay, it still only manages to produce around 35
polymers of maximum length. This is slightly more than for the reaction set that was evolved for
no-decay, but still significantly less than the more than 200 that can be reached without decay at
all. So, apparently the decay in this case is too high to maintain a large enough number of
maximum-length polymers, and the relevant ``information'' is lost or at least significantly reduced.

\begin{figure}
  \begin{center}
  \begin{tabular}{cc}
  \epsfig{file=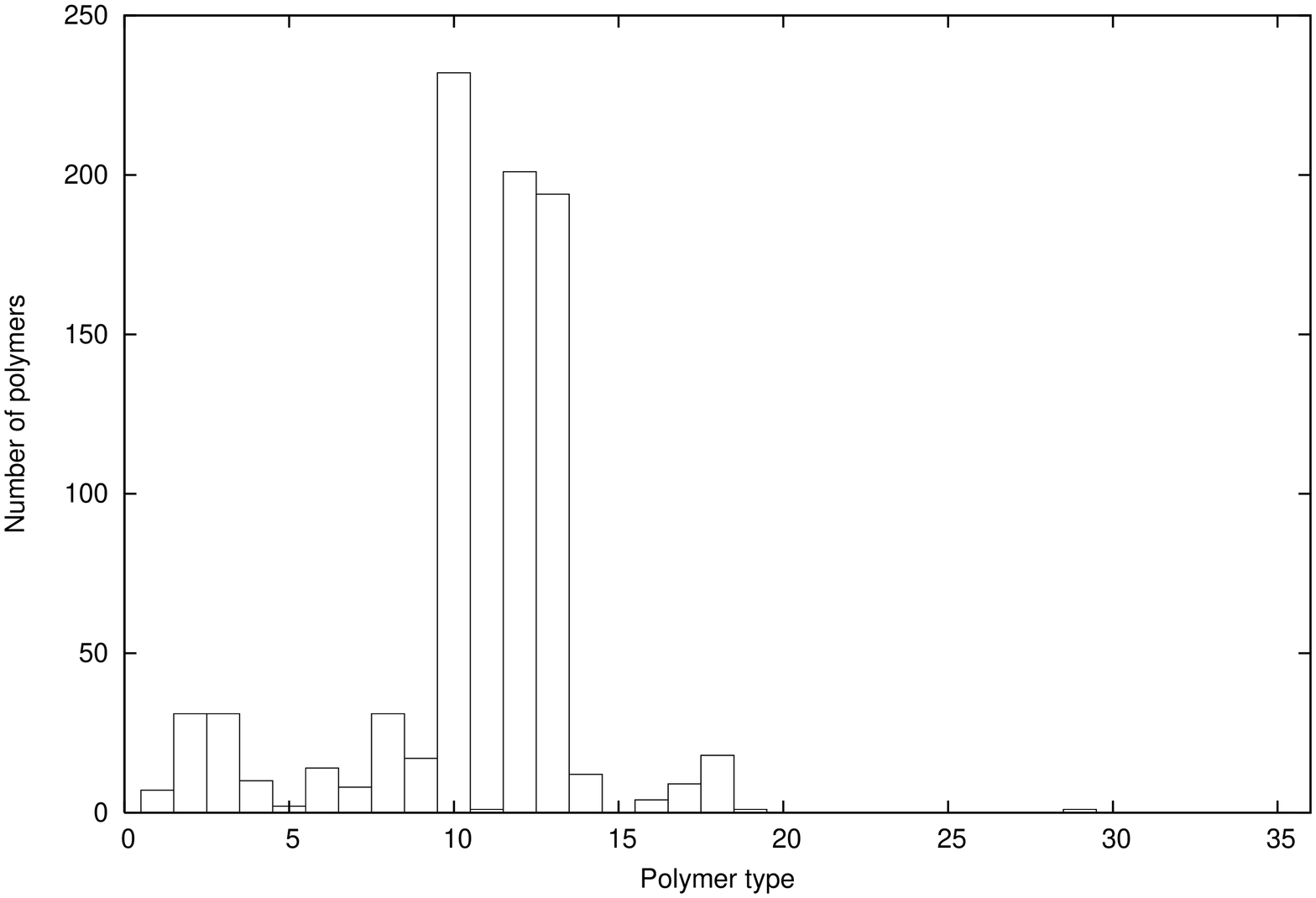, width=2in, angle=270} &
  \epsfig{file=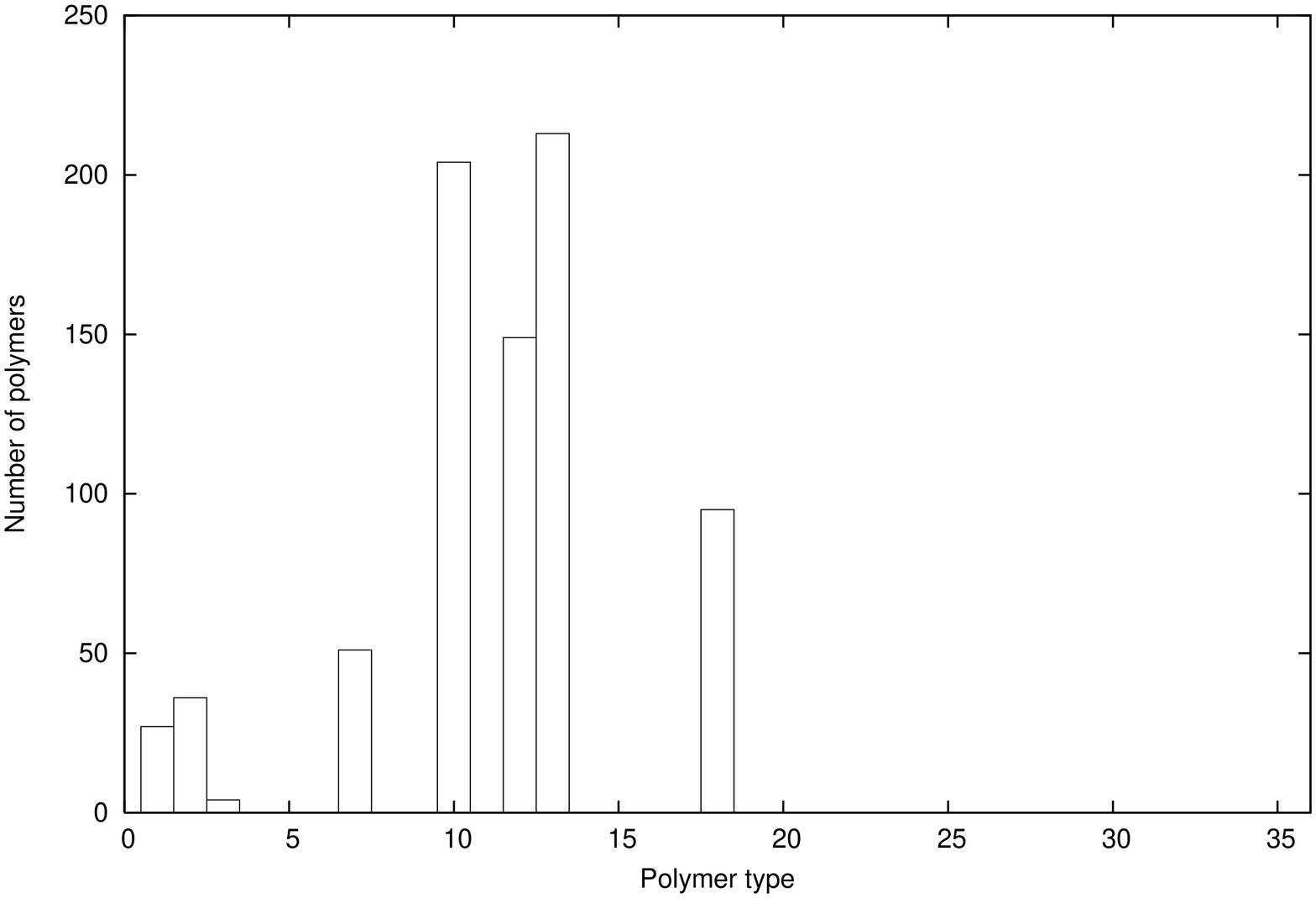, width=2in, angle=270}
  \end{tabular}
  \end{center}
  \caption{Left: The result for the best reaction set evolved for the second target (polymers of
           length 10, 12, and 13) with decay. This reaction set is able to maintain around 200 of
           each of the target polymers. Right: The results of the same reaction set, but iterated
           without any decay.}
  \label{fig05}
\end{figure}

Next, the second target, polymers of lengths 10, 12, and 13, is investigated. The GA was run
several times on this target, with the decay parameter at $d=0.0001$. Figure \ref{fig05} (left)
shows one of the best reaction sets evolved for this target. As the plot shows, it manages to
produce around 200 polymers of each length, roughly equal to the amount of polymers of maximum
length that can be produced without decay. So, even though the relevant ``information'' cannot be
maintained in one long polymer under the influence of decay, it can be maintained by dividing the
information up over smaller polymers. The information can be maintained at a similar level (around
200 polymers) using these smaller polymers.

It turns out that the performance of this particular reaction set is slightly less when iterated
without decay. Figure \ref{fig05} (right) shows the equilibrium distribution in this case. This
particular reaction set relies partly on decay to break down longer polymers into smaller ones,
which it can then use to create the target polymers. Without this breakdown, there are fewer
smaller polymers available to create the targets ones, resulting in a somewhat lower production of
target polymers.

\begin{figure}
  \begin{center}
  \epsfig{file=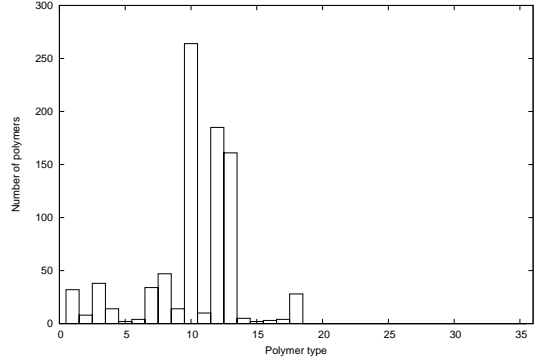, width=2in, angle=270}
  \end{center}
  \caption{The polymer distribution generated by the core set of 13 reactions (see table
           \ref{tab:core}) of the best reaction set.}
  \label{fig07}
\end{figure}

On investigating the evolved reaction set, it turns out that there is a core set of only 13
reactions (out of the 100), that are mainly responsible for its performance. When isolating these
13 reactions, and iterating this core set on the same initial polymer distribution (see figure
\ref{fig01}) and the same decay rate ($d=0.0001$), the equilibrium distribution is as shown in
figure \ref{fig07}. The total number of target polymers produced is slightly less than with the
complete set of 100 reactions, but it is still around 200 each. So, the other 87 reactions only
slightly increase the performance of this core set.

\begin{figure}
  \begin{center}
  \epsfig{file=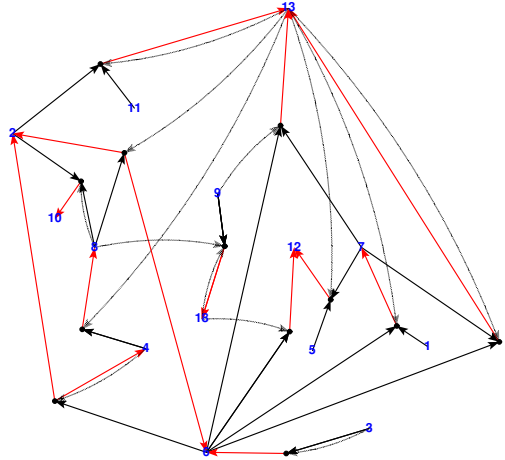, width=3in}
  \end{center}
  \caption{The reaction graph of the core set of 13 reactions (see table \ref{tab:core}).}
  \label{fig:core}
\end{figure}

Figure \ref{fig:core} shows the reaction graph of this core set of 13 reactions. The numbers
indicate the polymers types (or lengths) and dots indicate reactions. The black arrows going from
polymers to reactions indicate the reactants going into the reaction, and the red arrows going from
reactions to polymers indicate the products coming out of the reaction. The gray arrows indicate
the catalysts of each reaction. Table \ref{tab:core} lists the 13 reactions.

\begin{table}
  \begin{center}
  \begin{tabular}{|rrl|rrl|}
  \hline
  2+8+8   & $\rightarrow$ & 10+8  & 8+13   & $\rightarrow$ & 2+6+13 \\
  5+7+13  & $\rightarrow$ & 12+13 & 3+3+3  & $\rightarrow$ & 6+3    \\
  6+6+18  & $\rightarrow$ & 12+18 & 1+6+13 & $\rightarrow$ & 7+13   \\
  6+7+9   & $\rightarrow$ & 13+9  & 6+4    & $\rightarrow$ & 2+4+4  \\
  6+7+13  & $\rightarrow$ & 13+13 & 9+9+8  & $\rightarrow$ & 18+8   \\
  2+11+13 & $\rightarrow$ & 13+13 & 9+9+18 & $\rightarrow$ & 18+18  \\
  4+4+13  & $\rightarrow$ & 8+13  &        &               &        \\
  \hline
  \end{tabular}
  \end{center}
  \caption{The 13 reactions that form the core set of the evolved reaction set.}
  \label{tab:core}
\end{table}

As can be seen in the reaction graph, there are various ``hypercycle-like'' structures. For
example, polymer type 13, one of the targets, serves as a catalyst in 6 different reactions, 3 of
which produce target polymers. There are also several autocatalytic reactions, where the reaction
product catalyzes its own creation (such as in $6+7+13 \rightarrow 13+13$, and $9+9+18 \rightarrow
18+18$). Furthermore, there are several closed loops in the graph, where the polymer types in this
loop act alternately as reactants or catalysts and products. For example, $4+4+13 \rightarrow 8+13$,
$8+13 \rightarrow 2+6+13$, and $6+4 \rightarrow 2+4+4$ is such a loop, and there are several more.

One other thing to note is that some polymer types in this reaction graph are not directly produced
by one or more of the 13 reactions in the core set. For example, polymer types 11, 9, 4, 2 and some
others are only used as reactants or catalysts. However, the core set relies on decay to produce
these polymer types, by for example breaking down a polymer of length 13 into polymers of lengths
11 and 2, or 9 and 4, etc. So, instead of being hindered by decay, this reaction set has adapted to
actually make good use of the existence of decay!

Results on other GA runs were similar, but often with slightly lower performances of the evolved
reactions sets, or somewhat larger core sets. The result shown here was the best one found among
the different runs.

%
%
\section{Conclusions} \label{sec:Conclusions}

The amount of information that can be maintained in a population of reacting polymers depends on the
reaction efficiencies and the decay rate. For example, above a certain decay rate, it seems not
possible anymore to maintain a significant number of large polymers. However, as is shown here, it
is possible to evolve reaction sets that are able to maintain the relevant information by using a
set of smaller polymers, each of which holds only part of the information (in our case, the lengths
of the smaller polymers add up to the length of the large one, but one can imagine encoding
information in different and more sophisticated ways in polymers of different types and lengths).

So, whereas maintaining a certain amount of information in one large polymer breaks down at a
certain decay rate, splitting the information up over several smaller polymers makes it possible to
maintain the same amount of information (around 200 polymers of each type, in our case). In fact,
the evolved reaction sets actually learn to make use of the decay by eliminating the use of
reactions that create smaller polymers that can be used in building up the target ones. These
evolved reaction sets rely on the decay to create these smaller polymers. This gives rise to
relatively small core sets of reactions that are highly efficient and sufficient to reach the
desired target polymer distribution.

Moreover, in these core sets the beginnings of hypercycle-type structures can be observed in the
form of target polymers acting as catalysts, the existence of autocatalytic reactions, and several
closed loops in the reaction graph. These results can also have important indications for other,
more general questions relating to, e.g., the origin of life, where it is believed that
hypercycle-type structures were an important step in achieving the complexity necessary to support
living cells. The results presented here clearly show that it is indeed possible to evolve
hypercycle-type structures to maintain a certain amount of information under the influence of decay,
or polymer breakdown. This paper mainly present work in progress, but out results are very
encouraging, and demand further investigation into this phenomenon.

%
%
\section*{Acknowledgments}

This research was supported by FAPESP under project number 02/01038-7.

%
%
\bibliography{rsets}
\bibliographystyle{alpha}

\end{document}